\documentclass[twocolumn,prb,superscriptaddress,showpacs,citeautoscript,floatfix]{revtex4}

\usepackage{graphicx}
\usepackage{dcolumn}   % column centering
\usepackage{bm}        % bold math
\usepackage{flafter}
\usepackage{color}
\usepackage{hyperref}

\begin{document}
\title{High pressure structural and magneto-transport studies on type-II Dirac semimetal candidate Ir$_2$In$_8$S: Emergence of superconductivity upon decompression}

\author{Pallavi Malavi}
\email[E-mail:~]{spallavi@barc.gov.in}
 \affiliation{High Pressure and Synchrotron Radiation Physics Division, Bhabha Atomic Research Centre, Trombay, Mumbai 400085, India}

\author{Prakash Kumar}
 \affiliation{Department of Physics, Indian Institute of Science Education and Research (IISER), Pune 411008, India}

\author{Navita Jakhar}
 \affiliation{Department of Physics, Indian Institute of Science Education and Research (IISER), Pune 411008, India}

\author{Surjeet Singh}
\affiliation{Department of Physics, Indian Institute of Science Education and Research (IISER), Pune 411008, India}

\author{S. Karmakar}
 \affiliation{High Pressure and Synchrotron Radiation Physics Division, Bhabha Atomic Research Centre, Trombay, Mumbai 400085, India}

\date{\today}

\begin{abstract}
The structural and magneto-transport properties of type-II Dirac
semimetal candidate Ir$_2$In$_8$S have been investigated under high
pressure. The ambient tetragonal structure ($P4_2/mnm$) is found to
be stable up to $\sim$7 GPa, above which the system takes an
orthorhombic $Pnnm$ structure, possibly destroying the Dirac cones
due to the loss of the four-fold screw symmetry. In the tetragonal
structure, a gradual suppression of the transverse
magneto-resistance and a rapid change in the magnetic field
dependence above 50K suggest possible $T$-dependent Fermi surface
modification. In the high pressure phase, the metallic character
increases marginally (as evident from the increased RRR value)
accompanied with suppressed magneto-resistance, without emergence of
superconductivity up to 20 GPa and down to 1.4K. Most surprisingly,
upon release of pressure to 0.2 GPa, a sharp resistance drop below
$\sim$4K is observed, field varying measurements confirm this as the
onset of superconductivity. The observed changes of the carrier
density and mobility in the pressure-released tetragonal phase
indicate electronic structural modification resulting from the
irreversible polyhedral distortion. A simultaneous increase in the
residual resistivity and carrier density upon decompression
indicates that an enhanced impurity scattering play a key role in
the emergence of superconductivity in the tetragonal Ir$_2$In$_8$S,
making it an ideal platform to study topological superconductivity.

%Magneto-transport measurements give evidence of irreversible changes
%of the carrier density and mobility in the pressure-released
%tetragonal phase, indicating an irreversible electronic structural
%modification.
%Single crystal Ir$_2$In$_8$S has been investigated by resistance,
%magneto-resistance and Hall measurements under high P

%With the increasing quasi-hydrostatic pressures, the onset temperature of
%the local structural disorder (T*$\sim$230K) in the tetragonal phase
%systematically decreases and eventually disappears in the high
%pressure phase.

\end{abstract}

%\pacs{PACS: 74.70.Ad, 74.62.Fj, 71.45.Lr, 74.25.Dw}

%\pacs{71.45.Lr, %Dirac semimetal
%     74.25.Dw, %superconductivity
%     74.25.Ha, %High Pressure Structural Transition
%     71.20.Gj, Electronic structure of Semimetals,
%     74.62.Fj, Pressure effects on SC transition temperature variation,
%     }%

\maketitle

\section{Introduction}

Three dimensional Dirac semimetals (DSM) are new states of
topological quantum matter, characterized by symmetry-protected
linear band crossing at the 4-fold degenerate Dirac point near the
Fermi level~\cite{Young2012,Wang2018}. The massless Dirac fermions
arising from Dirac points are often considered responsible for the
remarkable transport properties, including ultrahigh carrier
mobility as a result of topology-protected suppression of
backscattering and extremely large and unusual field and angle
dependent magneto-resistance (MR) in 3D materials like
Cd$_3$As$_2$~\cite{Neupane2014,Liang2015},
Na$_3$Bi~\cite{Xiong2015,Xiong2016}, TlBiSeS ~\cite{Novak2015} and
PtBi$_2$~\cite{Gao2017}. Weyl semimetals (WSM) are topological
semimetals where either inversion symmetry or the time reversal
symmetry breaking causes splitting of the Dirac point into chiral
Weyl nodes having different surface states and Fermi arcs than
DSM~\cite{Xu2015}. Nonmagnetic WSMs in the TaAs-family, hosting Weyl
fermions and featuring chiral magnetic anomaly, also exhibit
ultrahigh mobility and extreme magnetoresistance
(XMR)~\cite{Huang2015,Shekhar2015,Zhang2015,Ghimire2015}. The XMR
has been reported in the Lorentz invariance-broken (type-II) WSM
candidates WTe$_2$, WP$_2$,
MoTe$_2$~\cite{Ali2014,Thoutam2015,Wang2017,Schoenemann2017,Chen2016}
and in Dirac line-node
semimetals~\cite{Wang2016a,Ali2016,Singha2017,Mun2012}.
Understanding the remarkable transport properties in terms of
non-trivial band topology is often debated. While electron-hole
compensation mechanism is considered a necessary ingredient for the
emergence of the XMR in multiband
semimetals~\cite{Ali2014,Wang2015,Zeng2016,Sun2016,Pavlosiuk2018},
the possible origin of the ultrahigh mobility (the most essential
criteria) is discussed in terms of either topological
protection~\cite{Liang2015,Xiong2015,Shekhar2015,Wang2016b,Zhang2017}
or spin-orbit coupling (SOC)-induced orbital texture on Fermi
surface~\cite{Jiang2015,Wang2017,Niu2016,Tafti2016,Sun2016}.

Lorentz invariance-broken (type-II) DSM state, as has been predicted
and experimentally verified in the
1$T$-PtSe$_2$-family~\cite{Huang2016,Yan2017} and
VAl$_3$-family~\cite{Chang2017}, has become of tremendous current
research interest due to the observed finite density of states
around the Dirac point, that may help emerge topological
superconductivity in the surface states and host Majorana modes with
possible applications in quantum computations~\cite{Alicea2011}. The
non-trivial topology has indeed been established in the
superconducting (SC) phases of PdTe$_2$ and
Ir$_{1-x}$Pt$_x$Te$_2$~\cite{Fei2017,Clark2018,Fei2018}. However,
the normal state MR and so the carrier mobility are found much
reduced in these compounds, implying non-involvement of the massless
Dirac fermions in the transport phenomena. It is noteworthy that the
observed SC in the type-I DSM Cd$_3$As$_2$ at high pressure is
accompanied by a structural transition~\cite{He2016} and SC emerges
in WTe$_2$ at high pressure with complete suppression of
MR~\cite{Kang2015,Pan2015}, suggesting possible non-coexistence of
the SC state and the topological surface states. Tremendous efforts
are thus being continued in search of new and diverse type-II DSM
with remarkable transport properties, as possible platform to
explore topological superconductivity.

Subchalcogenide compounds with diverse structural motifs have
recently been found to be promising topological semimetal
candidates~\cite{Liu2018,Wang2018,Khoury2019,Khoury2020}. The
sub-valent metallic character in related compounds also leads to
competing electronic phases like charge density wave (CDW) and
superconductivity~\cite{Sakamoto2007,Sakamoto2008}. Recently,
type-II Dirac semimetal candidate subchalcogenides Ir$_2$In$_8$Q
(Q=S, Se, Te) have been grown successfully and are also found to
exhibit significantly large and anisotropic MR (qualifying the
ultrahigh mobility criteria)~\cite{Khoury2019,Khoury2020}.
Ir$_2$In$_8$Q compounds crystallize in tetragonal $P4_2/mnm$ space
group with 3D framework of IrIn$_8$ polyhedra with chalcogen atoms
in the channels along the c-axis. No superconductivity has been
reported in this series of compounds. But an increased chalcogen
atom radii leads to lattice instability with commensurately
modulated structures due to enhanced polyhedral distortions at
intermediate temperatures. This also reduces the low temperature
magnetoresistance~\cite{Khoury2020}. As the effect of pressure is
expected to be similar to that of decreasing atomic radii, high
pressure study on these compounds may reveal competing electronic
phases, including superconductivity. Also the band structure
calculations show that the two sets of Dirac points in Ir$_2$In$_8$S
are situated $\sim$25 and $\sim$40 meV above the Fermi
level~\cite{Khoury2019}. Pressure-induced band broadening may shift
the Fermi level upward by enhancing the carrier density and help
tune the Fermi level towards the Dirac points, thus possibly
exhibiting exotic transport signatures.

We report here the effect of pressure on the structural and
magneto-transport properties of subsulfide Ir$_2$In$_8$S. The
ambient tetragonal structure ($P4_2/mnm$) is stable up to $\sim$7
GPa, above which the system takes an orthorhombic $Pnnm$ structure.
A loss of the four-fold screw symmetry in the high $P$ structure
possibly destroys the Dirac cones making it a trivial semimetal. In
the tetragonal structure, a drastically suppressed MR at high $T$
and the change in its field dependence suggests significant Fermi
surface (FS) modification with increasing temperature. The
characteristic temperature $T^*$, below which local polyhedral
distortion ceases, decreases upon increasing $P$. In the high $P$
orthorhombic phase, the metallicity increases marginally and the
magneto-resistance gets suppressed. No superconductivity has been
observed down to 1.4K up to $\sim$20 GPa. Surprisingly, upon
decompression to 0.2 GPa a sharp resistance drop is noticed below
$\sim$4K, indicating superconductivity onset in the
pressure-released sample. The magneto-transport properties show
significant change of the carrier density and mobility, suggesting
electronic structural modification due to subtle polyhedral
rearrangement upon $P$-cycling. An enhanced impurity scattering may
have an important role in the emergence of SC in the ambient
tetragonal structure (that supports type-II Dirac semimetal state),
making tetragonal Ir$_2$In$_8$S an ideal platform to study
topological superconductivity at ambient pressure.

%The present work will call for further investigations on P-released sample
%by surface probe techniques for getting direct evidence of Dirac physics (surface state)
%as well as topological nature of SC

\section{Experimental Methods}

High quality single crystal Ir$_2$In$_8$S has been synthesized using
indium metal flux method~\cite{Khoury2020} and characterized by
x-ray diffraction, EDX and HRTEM measurements (discussed in
Supplementary Material~\cite{misc1}). High-pressure powder x-ray
diffraction measurements at room temperature have been performed at
the XPRESS beam line ($\lambda=0.4957 {\AA}$) of the Elettra
synchrotron, Trieste. Single crystals were finely powdered and
loaded in a diamond anvil cell (DAC) for measurements under
quasi-hydrostatic pressures with methanol-ethanol-water (MEW)
(16:3:1) as pressure transmitting medium (PTM) and Cu as x-ray
pressure marker~\cite{Dewaele2004}. The 2D diffraction images were
recorded on a Mar345 image plate detector and these were converted
to $I-2\theta$ diffraction profiles using the Fit2D
software\cite{Hammersley1998}. Structural analyses were performed
using EXPGUI software~\cite{Larson1986}.

High-pressure resistance measurements at low temperatures have been
performed on a freshly cleaved $\sim$20 $\mu$m thick microcrystal of
$\sim$100 $\mu$m lateral dimension (cleaved plane have arbitrary
crystallographic orientation). A Stuttgart version DAC was used for
measurements under quasi-hydrostatic pressures up to 20 GPa using
NaCl as the PTM. The resistance was measured using standard four
probe method with 1 mA current excitation and in the ac lock-in
detection technique. For measurements down to 1.4 K, the DAC was
placed inside a KONTI-IT (Cryovac) cryostat. A nonmagnetic Cu-Be DAC
(M/s Easylab) was prepared with identical sample size for high-field
measurements under quasi-hydrostatic pressures up to 7.5 GPa and was
inserted into a S700X SQUID magnetometer (M/s Cryogenic Ltd) to
study transverse MR and Hall resistance up to 7T field. Pressures
were measured by conventional ruby luminescence method.

\section{Results and Discussion}

%\subsection{X-ray Diffraction}

\begin{figure}[tb]
\centerline{\includegraphics[width=90mm,clip]{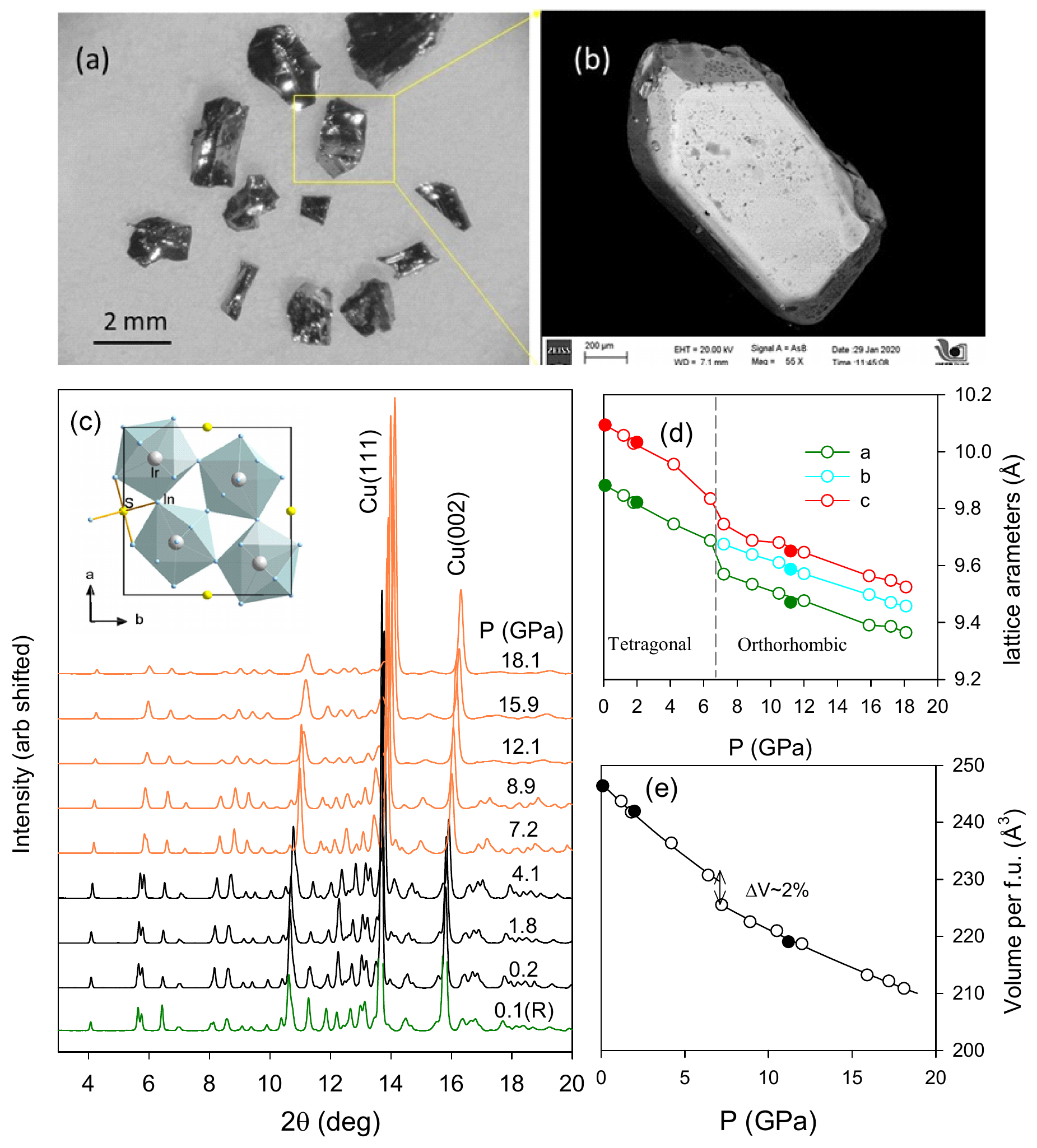}}
\caption{\label{Fig1} (Color online) (a) A few representative single
crystal specimens of Ir$_2$In$_8$S. (b) FESEM image of one of the
specimens. (c) Powder x-ray diffraction patterns at various high
pressures. Patterns in red correspond to the orthorhombic structures
above 7 GPa. Released pattern at 0.1 GPa is shown in green. Inset
shows the polyhedral arrangement in the tetragonal unit cell.
Variation of lattice parameters (d) and volume per formula unit (e)
are plotted as a function of $P$, solid circles are for releasing
pressures.}
\end{figure}

Powdered diffraction patterns at various high pressures are shown in
Figure 1a. The ambient tetragonal structure (SG: $P4_2/mnm$) of
Ir$_2$In$_8$S is found to be stable up to $\sim$7 GPa. Structural
analyses at various high pressures have been performed by Le-Bail
profile fitting~\cite{misc1}, using the reported atomic
coordinates~\cite{Khoury2020}. For pressures above 7GPa, the system
undergoes a subtle structural transition to a low symmetric
orthorhombic structure ($Pnnm$), as determined by the group-subgroup
analysis. In Figure 1(b,c) are shown the variation of lattice
parameters and the unit cell volume as a function of pressure,
showing $\sim$2\% volume collapse across the structural transition
(indicating first order nature of the phase transition). $P-V$ data
when fitted with the 3rd order Birch-Murnaghan equation of state
gives bulk moduli 85(4) GPa ($B^\prime$=4.1) and 104(5) GPa
($B^\prime$=5) in the tetragonal and orthorhombic phases
respectively. Due to our inability to refine all the In atom
positions in the structural analysis of the XRD patterns at high
pressures, we cannot comment on the pressure evolution of the
IrIn$_8$ polyhedral distortions. Upon release of pressure, the
orthorhombic phase transforms back completely to the ambient
tetragonal structure, also regaining the initial Bragg peak width
(indicating absence of any apparent structural
disorder)~\cite{misc1}.

\begin{figure}[tb]
\centerline{\includegraphics[width=90mm,clip]{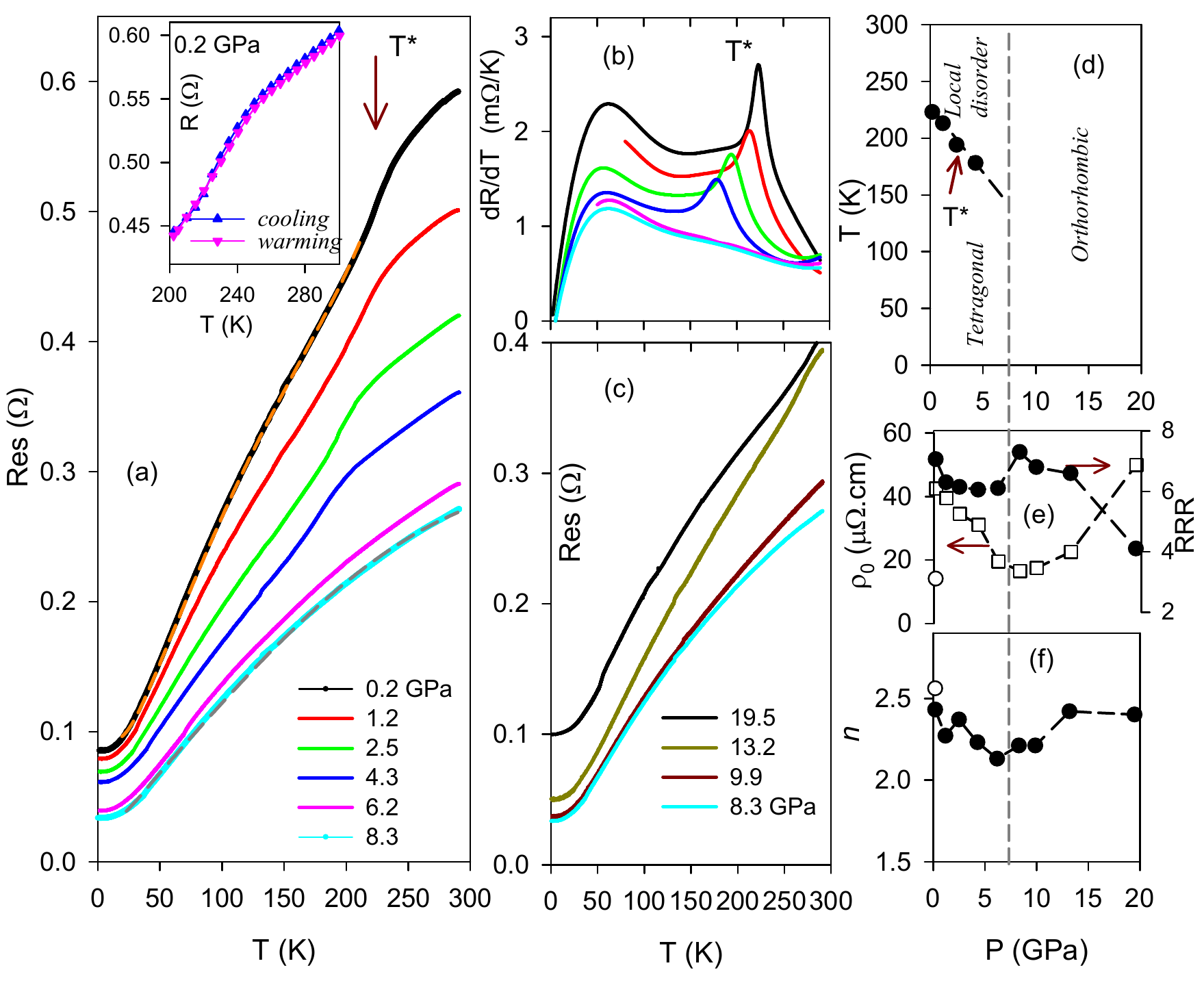}}
\caption{\label{Fig1} (Color online) (a) $R(T)$ plots at various
high pressures up to 8.3 GPa. Dashed lines are the BG fit for the
$R(T)$ data at 0.2 GPa and 8.3 GPa. Inset, $R(T)$ behavior near the
onset of local disorder in cooling and warming cycle. (b) $dR/dT$
plots showing systematic decrease of $T^*$. (c) $R(T)$ plots at
higher pressures up to 19.5 GPa (d) $P-T$ structural phase diagram
of Ir$_2$In$_8$S. (e,f) Pressure variation of the residual
resistivity, the RRR value and the exponent $n$ in the power-law fit
of $R(T)$ below 20K.}
\end{figure}

High pressure resistance data up to $\sim$20 GPa measured on the
single crystal Ir$_2$In$_8$S (along an arbitrary crystallographic
plane) in the temperature range 1.4-300K are shown in Figure 2(a,c).
No superconductivity has been observed down to 1.4 K. At 0.2 GPa,
$R(T)$ displays an overall metallic behavior, with anomalous kink
feature at the characteristic temperature ($T^*\sim$230K), above
which local disordering in the IrIn$_8$ polyhedra emerges, agreeing
well with the previous reported data~\cite{Khoury2020}. The presence
of noticeable hysteresis at high $T$ (shown as inset in Figure 2a)
supports the structural origin of this resistance anomaly. In the
ordered phase (below $T^*$) the resistance $R(T)$ above 30K follows
the Bloch-Gr\"{u}neisen (BG) resistivity model ~\cite{Ziman1962}
$$\rho(T)=\rho_0+C(\frac{T}{\Theta_D})^{k}\int_{0}^{{\Theta_D}/T}\frac{x^k}{(e^x-1)(1-e^{-x})}dx$$
where $\rho_0$ is the residual resistivity and $\Theta_D$ is the
Debye temperature. The $R(T)$ data for pressures up to 5 GPa can be
fit well with k=5, indicating phonon dominated scattering mechanism.
At 0.2 GPa, $\Theta_D$=214K , in reasonable agreement with the
reported value~\cite{Khoury2020}. The $R(T)$ data above 30K at 8 GPa
(in the orthorhombic phase) when fitted with the BG equation gives
k=4.1 with $\Theta_D$=120K, indicating significant electronic
structural modifications.

In Figure 2b are shown the $dR/dT$ plots at various pressures. The
characteristic temperature ($T^*$) down to which the random
polyhedral disorder persists in the tetragonal structure, as
indicated by the peak position, systematically decreases due to
enhanced intrinsic disorder in the system at higher pressures. In
the high pressure orthorhombic phase, the resistance anomaly
disappears indicating ordered polyhedral arrangements. Figure 2d
summarizes the structural phase diagram of Ir$_2$In$_8$S, based on
the XRD and resistance measurements. At a low pressure (0.2 GPa),
$R(T)$ data below 20K, fitted with power law $\rho(T)=\rho_0+A_1T^n$
gives $n\sim$2.5 indicating electron-electron dominated scattering
along with interband electron-phonon scattering~\cite{Ziman1962}.
Although the residual resistivity $\rho_0$ in the tetragonal
structure decreases significantly with pressure, the residual
resistivity ratio (RRR) and so the metallic character remains mostly
unchanged up to 5 GPa. An order of magnitude less RRR value in our
DAC-based measurement, compared to the reported value on the bare
sample~\cite{Khoury2019} (that were measured in the ab-plane), can
be attributed to the arbitrary crystallographic orientation of the
measurement plane. The metallicity (as seen from the RRR values)
marginally increases across the transition to the orthorhombic phase
but decreases more rapidly at higher pressures (Figure 2e). In the
tetragonal phase, $n$ decreases systematically with increasing $P$
and approaches the Fermi Liquid behavior ($n=2$) near the structural
transition, with opposite trend noticed in the orthorhombic phase
[Figure 2f]. The structural transition is thus associated with a
significant change in the electronic structure. Moreover, in the
high pressure orthorhombic structure ($Pnnm$), the four-fold screw
$4_2$ symmetry (that protects Dirac point linear band crossing in
the tetragonal structure~\cite{Khoury2020}) is lost, suggesting a
possible quantum topological transition associated with the
structural transition.

\begin{figure}[tb]
\centerline{\includegraphics[width=90mm,clip]{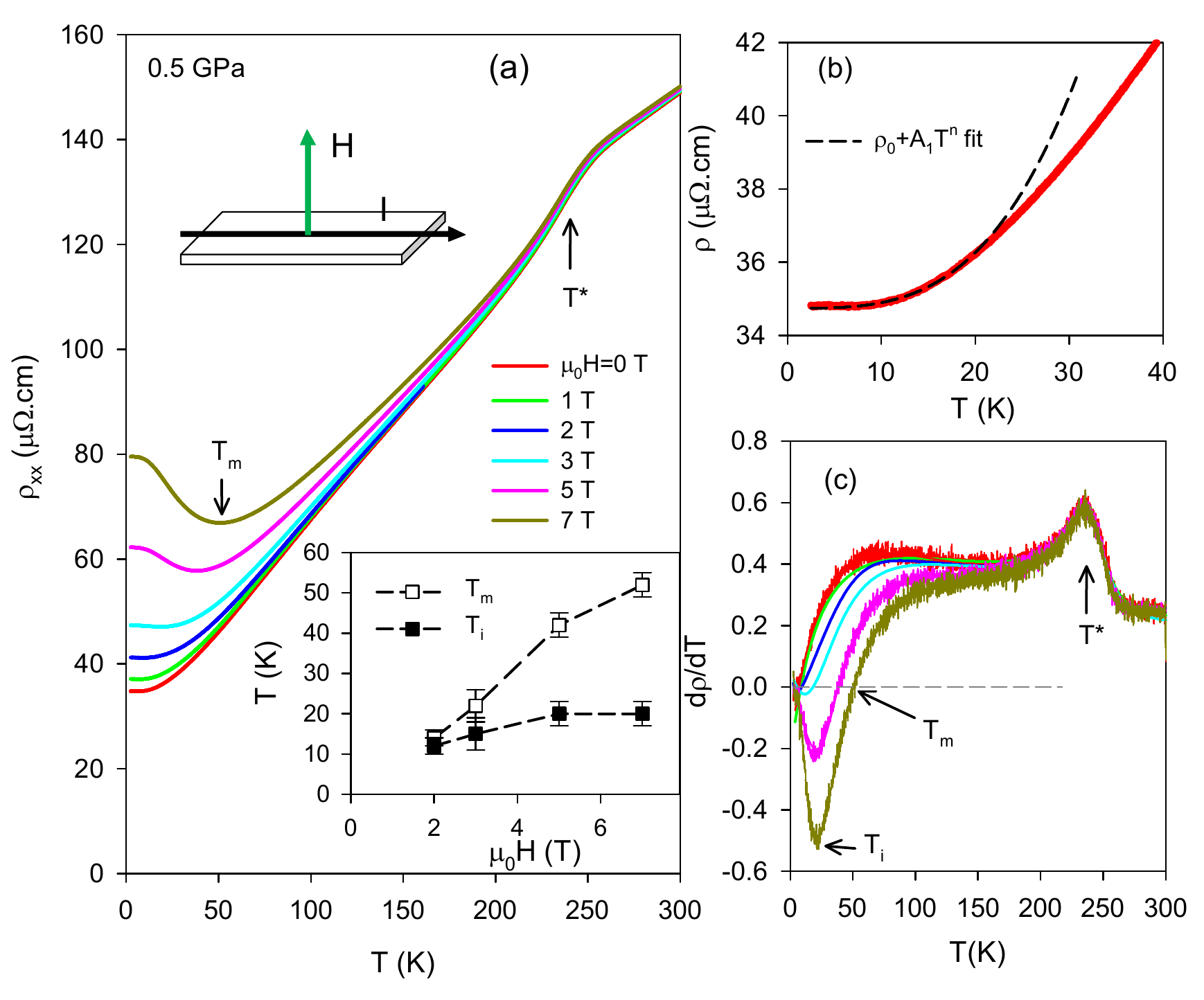}}
\caption{\label{Fig1} (Color online) (a) Resistivity $\rho_{xx}$
plotted as a function of temperature at 0.5 GPa under different
magnetic fields. Inset, The $T-H$ phase diagram, field variation of
the two characteristic temperatures $T_i$ and $T_m$, as obtained
from (c). (b) Zero field $\rho(T)$, obeying the power law fit below
20K. (c) $d\rho/dT$ plots at various fields.}
\end{figure}

First, we show the temperature dependence of resistivity in
Ir$_2$In$_8$S at a low pressure (0.5 GPa) in different magnetic
fields applied perpendicular to the current plane (see Figure 3a).
With the application of the magnetic field, $\rho(T)$ shows a large
upturn which saturates below $\sim$20 K. The $\rho(T)$ is also found
to obey the power law $T$-dependence below $\sim$20K (Figure 3b).
The signature is similar to that of known extreme magneto-resistance
materials~\cite{Mun2012,Takatsu2013,Wang2014,Ghimire2015,Ali2014,Thoutam2015,Schoenemann2017,Chen2016,Wang2017,Gao2017,Shekhar2015,Zhang2015,Tafti2016}.
Based on the field dependence of $d\rho/dT$ (Figure 3c) we plot the
temperature-field phase diagram as inset in Figure 3a, where $T_m$
and $T_i$ are taken as the sign change point and the minimum in
$d\rho/dT$~\cite{misc2}. The phase diagram is also consistent with
known XMR materials~\cite{Tafti2016}. We can note that the
order-disorder transition temperature $T^*$ is independent of the
applied magnetic field up to 7T, further supporting its structural
origin.

\begin{figure}[tb]
\centerline{\includegraphics[width=90mm,clip]{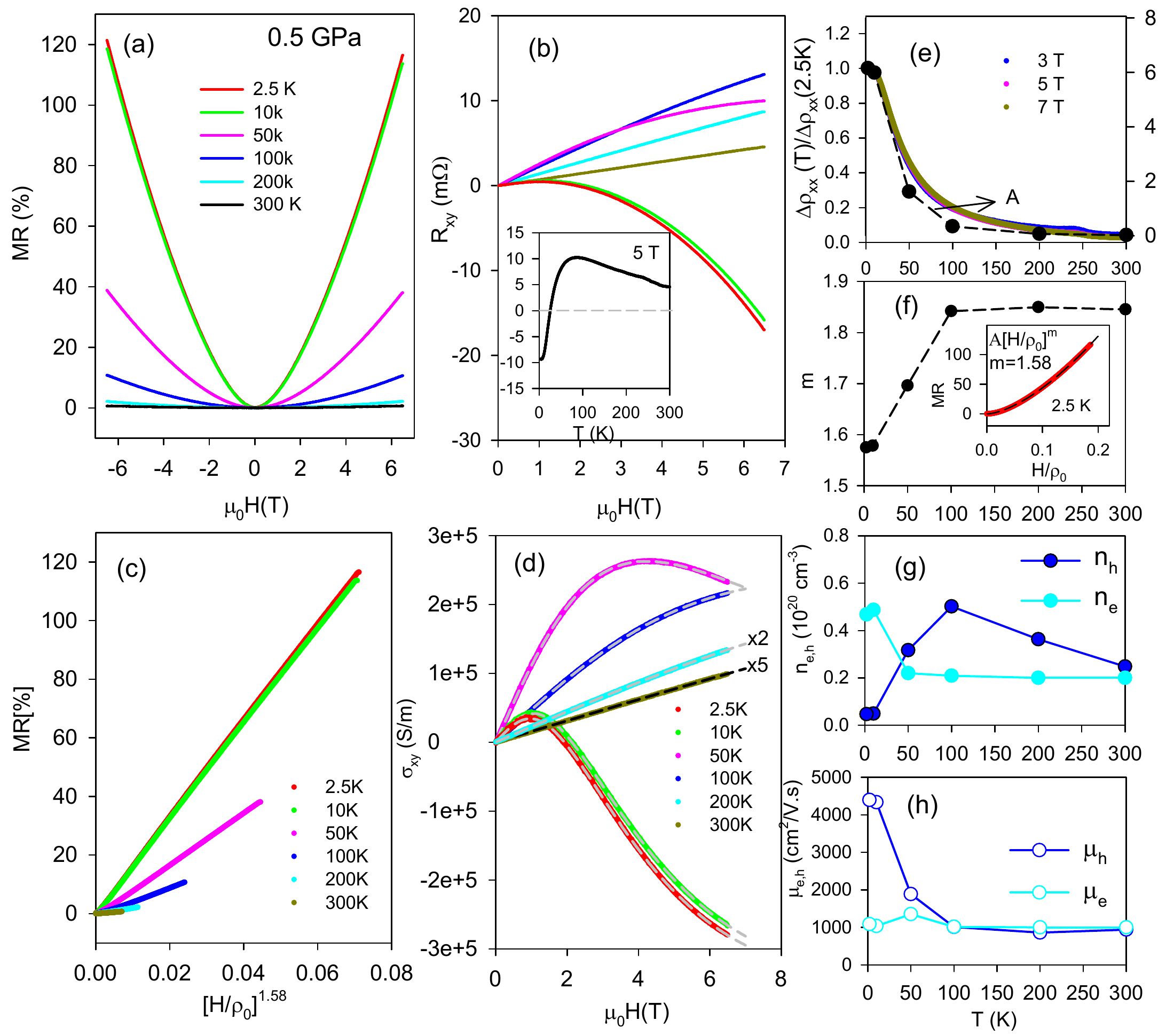}}
\caption{\label{Fig1} (Color online) Transverse MR (a) and Hall
resistance (b) at 0.5 GPa as a function of applied magnetic field at
various temperatures. Inset in (b), temperature dependent Hall
resistance at 5T field. (c) MR vs $[H/\rho_0]^{1.58}$ plots at
different $T$. Above 10 K, curves do nor superpose, showing
violation of Kohler's rule. (d) Plots of Hall conductivity
$\sigma_{xy}$. (e) Field-induced normalized resistivity change
plotted as a function of $T$. The A coefficient is found to scale
with the resistivity change (f) $T$-variation of the exponent $m$.
(g,h) $T$-variation of the carrier density $n_{e,h}$ and mobility
$\mu_{e,h}$, as obtained from the Hall conductivity from the
two-band model fit.}
\end{figure}

Figure 4a,b show the plots of the field dependence of the transverse
magneto-resistance MR=[R(H)-R(0)]/R(0) and the Hall resistance at
various temperatures at 0.5 GPa. As the measurements on the flake
sample were performed in the van der Pauw method, the observed
asymmetry of the MR curves originates from the in-plane Hall
contribution that is separated by symmetrizing the MR curves,
MR$_{sym}$(H)= [MR(H)+(MR(-H)]/2. An order of magnitude reduced
transverse MR in our measurement, compared to the reported value for
the ambient sample~\cite{Khoury2019}, can be due to the arbitrary
crystallographic plane of this high $P$ measurements. Moreover, a
small pressure can lead to drastic reduction of MR in XMR materials
due to Fermi surface modification~\cite{Cai2015}. With increasing
temperature, MR decreases rapidly. A highly non-linear field
dependent Hall resistance indicates multicarrier transport behavior
in Ir$_2$In$_8$S (Figure 4b). At low temperature large negative
$R_{xy}$ indicates electron dominated transport. At high temperature
above 50K, $R_{xy}$ becomes positive overall, showing hole
dominating transport. This is also apparent in the $T$-dependent
Hall resistivity measured at 5T field (see inset of Figure 4b,
obtained by field reversal and current reversal measurements). The
presence of multiple Fermi surfaces in the tetragonal Ir$_2$In$_8$S
have been reported by quantum oscillation measurements and band
structure calculations~\cite{Khoury2019}.

Field-induced resistivity upturn in the XMR materials can be
explained by semiclassical multi-carrier model in systems obeying
modified Kohler's rule
$MR=F[H/\rho_0]=A[H/\rho_0]^m$~\cite{Wang2015,Sun2016}. In case of
perfect electron-hole resonance condition, $m$=2~\cite{Ali2014}. For
systems where Kohler's rule~\cite{Pippard1989} is nearly obeyed
(m$\approx$2 and $T$- independent), MR is found to scale with
$A$~\cite{Wang2017}. As shown in Figure 4c, MR in Ir$_2$In$_8$S
systematically deviates from Kohler's rule above 10K (both $A$ and
$m$ varying rapidly with $T$). At 2.5K, $m$=1.58 (see inset in
Figure 4f), similar to other type-II topological
semimetals~\cite{Wang2017,Pavlosiuk2018a}; deviation from the
quadratic field dependence ($m<$2) can be due to un-compensated
carriers or anisotropic Fermi surface or field-induced Fermi surface
modification~\cite{Li2018,Thoutam2015}. In case of anisotropic
multiband materials, relative contribution of different FS pockets
also vary with strength and orientation of the magnetic
field~\cite{Kumar2016,Han2017}. Variation of $A$ with temperature,
when compared with field-induced resistivity change (shown in Figure
4e), indicates a rapid change of carrier transport behavior below
50K. This is further verified by the observed variation of the
exponent $m$ as a function of temperature (Figure 4f) indicating
electronic structural modification at low $T$. The field dependent
MR thus follows MR$\propto(\mu_aH)^m$, where $\mu_a$ is the average
carrier mobility.

The Hall conductivity $\sigma_{xy}$ has been obtained using the
formula $\sigma_{xy}=[\rho_{xy}/(\rho_{xx}^2+\rho_{yx}^2)]$, because
of the observed isotropic in-plane linear resistivity ($\rho_{xx}$)
and Hall resistivity $\rho_{xy}$. The carrier density and mobility
have been calculated by analyzing the Hall conductivity using two
band model~\cite{Hurd1972}
$$\sigma_{xy}=eB(\frac{n_h\mu_h^2}{1+\mu_h^2B^2}-\frac{n_e\mu_e^2}{1+\mu_e^2B^2})$$
where $n_e$ ($n_h$) and $\mu_e$ ($\mu_h$) are electron (hole)
density and mobility respectively. Figure 4d shows the plots of Hall
conductance $\sigma_{xy}$ at various temperatures. As shown in
Figure 4(g,h),  at low $T$ the electron density is an order of
magnitude higher than the hole density and therefore dominate the
transport. But the hole mobility is 4 times higher than the electron
mobility at this $T$. Our results are different than the reported
results from the measurements on ab-plane in a bare
sample~\cite{Khoury2019}. The discrepancy may be attributed to the
fact that the sample is under pressure in the present case or
intrinsic to the sample quality. The large carrier mobility mismatch
and a rapid decrease of hole mobility with increasing $T$ are
responsible for the systematic violation of the Kohler's rule with
$T$.

\begin{figure}[tb]
\centerline{\includegraphics[width=90mm,clip]{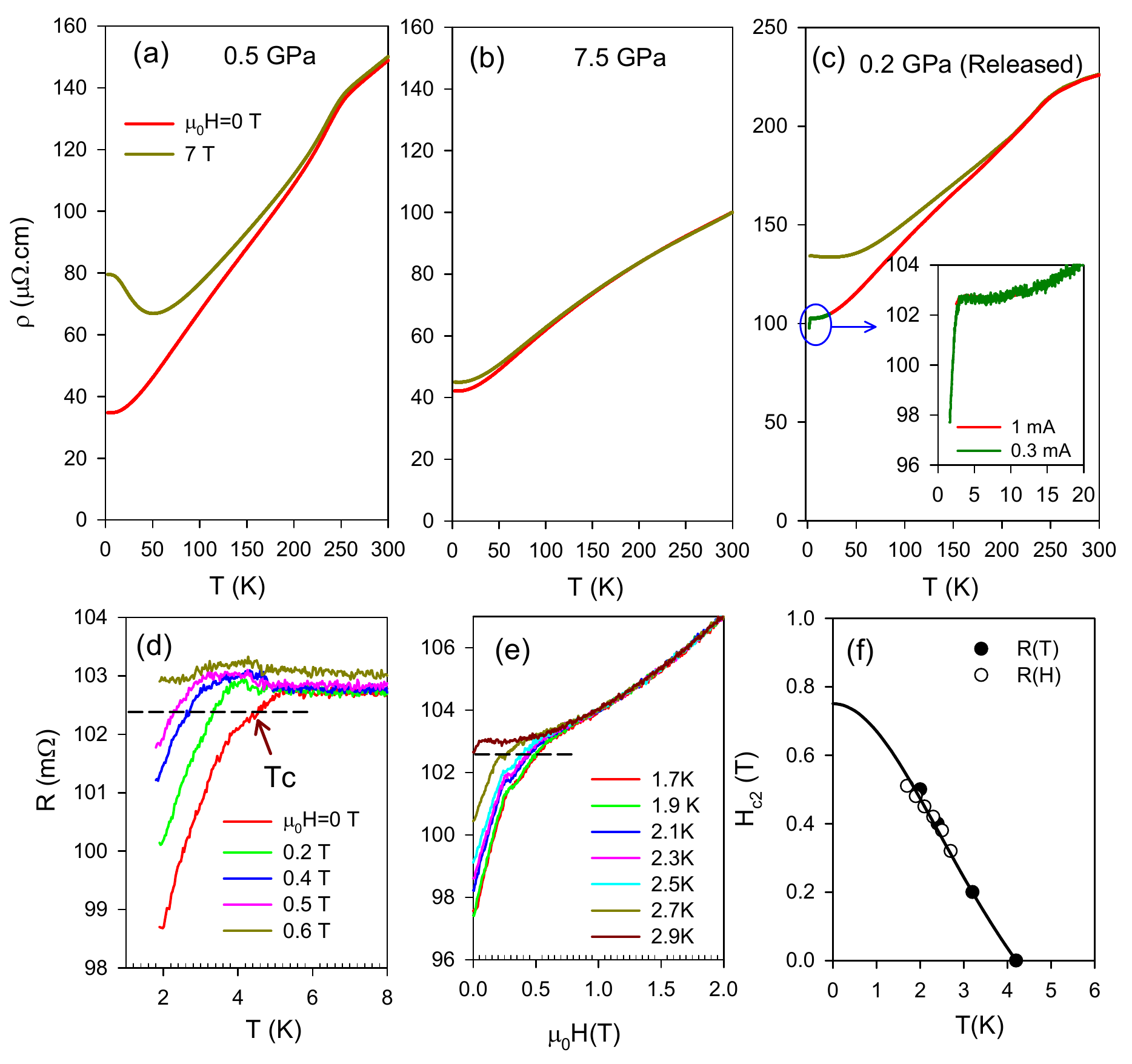}}
\caption{\label{Fig1} (Color online) Resistivity $\rho(T)$ at zero
field and at 7T field plotted at (a) 0.5 GPa, (b) 7.5 GPa and (c) at
0.2 GPa (upon decompression). Resistivity drop below $\sim$4K in (c)
is shown by blue circle, enlarged view shown as inset. (d) $R-T$
data near SC onset $T_c$ at 0.2 GPa (upon decompression) under
different fields up to 0.6 T. (e) $R-H$ data at various $T$ below SC
$T_c$ (f) the $H_{c2}$-$T_c$ plots and the Ginzburg-Landau fit.}
\end{figure}

Magneto-resistance (MR) and Hall measurements have been performed up
to 7T field and at pressures up to 7.5 GPa. Figures 5(a-c) show the
$\rho(T)$ plots for B=0T and 7T at 0.5 GPa, 7.5 GPa and at 0.2 GPa
(upon decompression). With increasing pressure, the field-induced
resistance upturn decreases with dramatic suppression of MR. Upon
decompression at 0.2 GPa, the residual resistivity at zero field
increases and MR is partially recovered (Figure 5c). To our
surprise, we observe a sharp resistance drop below $\sim$4K, that
becomes prominent at low current excitation, as can be seen in inset
in Figure 5c. This indicates possible onset of superconductivity in
the pressure-released sample. It is worth mentioning that any In
precipitation upon $P$-cycling causing the observed SC is unlikely,
as no additional Bragg peaks corresponding to elemental In have been
detected in XRD upon $P$-release~\cite{misc1}. Also the observed
broad transition width and much higher critical field (as shown
below) is in contrast to that of bulk In superconductivity. The
observed broad SC onset in $P$-released sample after repeated
$P$-cycling further verifies that SC occurs in the tetragonal
Ir$_2$In$_8$S and rules out the possibility of sample
decomposition~\cite{misc1}. Noting that Ir$_2$In$_8$S is prone to
disorder due to IrIn$_8$ polyhedral distortion, the observed
enhanced residual resistivity in the normal state may be attributed
to the increased defect/impurity scattering or the electronic
structural modification due to subtle structural rearrangements that
has not been detected by XRD. The impurity scattering induced
enhancement of $T_c$ has earlier been reported in In-doped
SnTe~\cite{Novak2013}. Also, SC persisting at lower $P$ with
enhanced $T_c$ in layered chalcogenides upon $P$-cycling has been
discussed in terms of structural
irreversibility~\cite{Ke2017,Dutta2018}. In the present study, no
intrinsic broadening of the XRD Bragg peaks has been observed upon
release of pressure, highlighting the absence of noticeable
structural disorder. Note that the resistance anomaly of local
disorder is regained with almost unchanged characteristic
temperature $T^*$. This indicates that SC emerges in the low
temperature disorder-free tetragonal structure in Ir$_2$In$_8$S.
Although the field-induced resistance change ($\Delta\rho$) is
nearly unchanged after $P$ release, a reduced XMR feature is
primarily due to the large residual resistivity. Although the MR at
5K and 7T field in the $P$-released sample is half of the initial
MR, the low $T$ resistance upturn with plateau (the XMR feature) is
still observed~\cite{misc1}. A reduced T$_m$ ($\sim$27K at 7T field)
is in agreement with the increased $T$-power law coefficient $A_1$,
as discussed by Sun et al.~\cite{Sun2016}.

These observations make $P$-released Ir$_2$In$_8$S a unique system,
where a large MR persists above SC $T_c$. As the SC transition width
is significantly broad, zero resistance is not reached down to 1.2K,
the lowest $T$ of this measurement. The onset T$_c$, taken as the
temperature with 1$\%$ resistance drops from the normal state (shown
by dashed line), decreases systematically with increasing magnetic
field (both in field-scanning and temperature-scanning mode, Figure
5d,e). The $T_c$-$H_{c2}$ plot [Figure 5f], when fitted with the
Ginzburg-Landau (GL) equation $H_{c2}=H_{c2}(0)[(1-t^2)/(1+t^2)]$
with t=$T/T_c$, estimates an upper critical field $H_{c2}(0)$
$\sim$0.75 T, which is an order smaller than the Pauli limit of
1.84$T_c\sim$7.5T. The estimated GL coherence length
($\xi_{GL}\sim$10nm) is much less than the transport mean free path
($l_m\sim$100 nm), suggesting the phonon mediated SC in the clean
limit. In case of orbital-limited behavior, predicted by
Werthamer-Helfand-Hohenberg (WHH) theory for the s-wave
superconductor~\cite{Werthamer1966}, the estimated upper critical
field $H_{c2}^{orb}\approx0.7
T_c\times(\frac{dH_{c2}}{dT})_{T=T_c}$=0.6 T. The measured upper
critical field thus exceeds the orbital-limited value. A
quasi-linear $H_{c2}-T_c$ plot can also be seen within our measured
$T$-range. Such quasi-linear SC behavior of Bi$_2$Se$_3$ at high
pressure has been ascribed to unconventional spin-orbit coupled
superconductivity~\cite{Kirshenbaum2013}. However, measurements at
further low temperatures are needed for better understanding of the
pair-breaking mechanism in the SC state of Ir$_2$In$_8$S.

We now focus on the results of magneto-resistance and Hall
measurements at various high pressures. Figure 6a shows the
transverse magneto-resistance plots at 2.5K, at various pressures up
to 7.5 GPa and at 5K upon $P$-release. The MR value decreases
rapidly with increasing pressure (see upper inset in Figure 6a). The
power-law field dependence ($MR\propto(\mu_a H)^m$, $m$=1.58 at 0.5
GPa) is found to change gradually at higher $P$ (see lower inset in
Figure 6a), showing $P$-induced change of the Fermi surface
anisotropy. In the orthorhombic phase (at 7.5 GPa), $m$ decreases
marginally. Upon release, at 0.2 GPa $m$ returns to the initial
value.

\begin{figure}[tb]
\centerline{\includegraphics[width=90mm,clip]{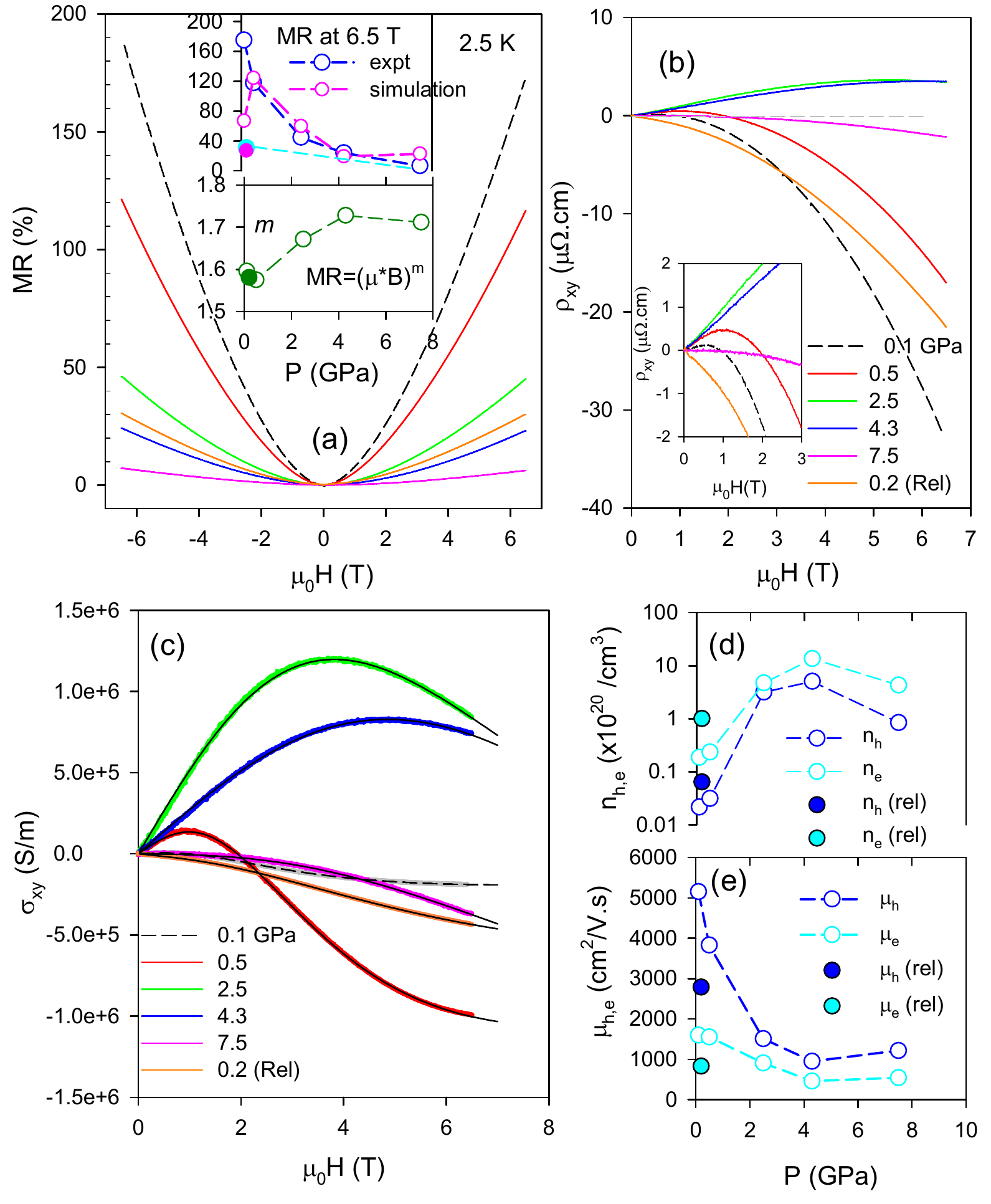}}
\caption{\label{Fig1} (Color online) (a) Transverse MR measured at
2.5 K at various quasi-hydrostatic pressures and upon decompression.
Upper inset, MR at 6.5 T field plotted as a function of pressure,
lower inset, the plot of the exponent for the field variation of the
MR as a function of pressure. (b) Hall resistivity plotted as a
function of magnetic field at various pressures. (c) Hall
conductivity plots $\sigma_{xy}(H)$ and the two-band model fit at
various high pressures (d,e) carrier density $n_{e,h}$ and mobility
$\mu_{e,h}$ as obtained from the Hall conductivity fit for different
pressures. In upper inset in (a), the simulated MR is also plotted
based on these carrier density and mobility values. The solid
circles in various plots correspond to the decompression data.}
\end{figure}

The field variations of the Hall resistivity at various pressures
are shown in Figure 6b. At low pressures, high field Hall
resistivity clearly indicate electron dominated transport, but
strong non-linearity and positive Hall resistivity at low field (see
inset in Figure 6b) indicates significant high mobility hole
contributions. To calculate the pressure dependence of the various
carrier density and their mobility, we have analyzed the field
dependence of the Hall conductivity $\sigma_{xy}$ using the two band
model fit (Figure 6c). In Figure 6(d,e) are plotted the pressure
variation of the carrier densities ($n_{e,h}$) and mobilities
($\mu_{e,h}$) at 2.5K. With increasing pressures in the tetragonal
phase, both carrier densities systematically increase. At low $P$,
the hole density is an order of magnitude less than the electron
density, but become roughly of same order above 2 GPa. The hole
mobility at low $P$ is a factor of 4 higher than electron mobility,
but decreases rapidly at 2 GPa where hole mobility becomes a factor
of 1.5 higher than the electron mobility. An enhanced field
dependence exponent m ($\sim$1.72) above 2 GPa can thus be
corroborated from the carrier density approaching the same order and
also systematic reduction of their mobility mismatch. The carrier
densities decrease rapidly in the orthorhombic phase (at 7.5 GPa).
Upon releasing $P$ to 0.2 GPa, the carrier densities remain at an
order of magnitude high value and the carrier mobility remains at a
relatively low value as compared to the initial $P$ values,
indicating the irreversible modification of the Fermi surface
pockets in the $P$-cycled sample.

In the two-band model the magneto-resistance is represented by the
equation~\cite{Pippard1989}:
$$MR=\frac{n_en_h\mu_e\mu_h(\mu_e+\mu_h)^2B^2}{(\mu_en_e+\mu_hn_h)^2+B^2\mu_e^2\mu_h^2(n_e-n_h)^2}$$ We have calculated the MR at 2.5K using the values of carrier
density and mobility obtained from the Hall conductivity fit and
have compared with the measured values (see upper inset of Figure
6a). Interestingly, the observed MR values at all pressures
(including that upon decompression), except at the lowest pressure
are in good agreement with the semi-classical two-band model. So the
Dirac points have little influence on the mobility at higher $P$. A
significant deviation at low pressure (exhibiting large MR and high
hole mobility) may have an origin beyond the classical description.
A topological origin or the SOC-coupled orbital texture may have
significant role in the enhanced transport mobility at low $P$ which
demand further theoretical and experimental investigations. Although
the MR (above SC $T_c$) in the $P$-released sample shows XMR-like
feature, its transport behavior follows the simple two-band model
suggesting involvement of trivial bands. However, as the system
returns to the initial crystal structure, we cannot rule out the
presence of Dirac like band crossings near the Fermi level even in
the presence of the enhanced impurity scattering. On the other hand,
the observed enhanced carrier density indicates SC onset may be
driven by the enhanced DOS at the Fermi level. This may also cause
shifting of the Fermi level towards the Dirac point, suggesting
possible coexistence of SC and Dirac cones. Our results thus call
for direct investigations on the $P$-released Ir$_2$In$_8$S by ARPES
measurements. As the zero-resistance has not been observed in SC
state in the $P$-released sample, further investigations at lower
$T$ will help understand its SC properties, especially if the SC is
of filamentary nature. Signature of unconventional nature of the
observed SC also suggests for point contact spectroscopic
investigation to probe for its possible origin in the topological
surface states.

\section{Conclusions}

The structural and magneto-transport properties of the type-II Dirac
semimetal candidate Ir$_2$In$_8$S have been investigated under high
pressure. The ambient tetragonal structure with four fold screw
symmetry that protects Dirac points is found to be stable up to
$\sim$7 GPa. In the low pressure phase, a significant change in
magneto-transport behavior with increasing $T$ suggests systematic
electronic structural modification. Surprisingly, upon release of
pressure a sharp resistance drop below $\sim$4K is observed, field
dependent studies on which verifies it as the onset of
superconductivity. High pressure magneto-transport measurements show
irreversible changes of the carrier density and mobility in the
pressure-released tetragonal phase, suggesting an irreversible
electronic structural modification. The enhanced impurity scattering
may have an important role in the emergence of superconductivity in
the ambient tetragonal structure (that hosts type-II Dirac semimetal
state), making it an ideal platform to study topological
superconductivity and Majorana physics. Present results will invite
investigations to explore possible emergence of SC in other
topological semimetals upon $P$-cycling by enhancing impurity
scattering in an irreversible manner, retaining the structural
symmetry and so the topological band structure.

\begin{acknowledgments}
High pressure XRD investigations have been performed at Elettra
synchrotron, Trieste under the proposal id 20195289. Financial
support from DST, Government of India is gratefully acknowledged.

\end{acknowledgments}

\bibliography{Ir2In8S}

\begin{thebibliography}{66}
\bibitem[{mis({\natexlab{b}})}]{misc2}
\bibinfo{note}{Resistivity upturn apperas in XMR materials beow $T_m=\hbar
  cB/m^*k_B$, where $m^*$ is the average effective mass. In Ir$_2$In$_8$S, $m^*
  \sim$ 0.3 $m_e$ and so $T_m\sim$ 50K at 7 T field.}
\end{thebibliography}
\end{document}